\documentclass[12pt]{article}
\usepackage[a4paper, top=0.8in, bottom=1in, left=1in, right=1in]{geometry}
\usepackage{amsmath, amssymb}
\usepackage{graphicx}
\usepackage{titlesec}
\usepackage{setspace}
\usepackage{url}
\usepackage{hyperref}
\usepackage{cite}
\usepackage{enumitem}
\usepackage{titling}
\usepackage{float}
\usepackage{caption}
\usepackage{xcolor}
\usepackage{indentfirst}
\usepackage{booktabs}
\usepackage{multirow}
\setlist{noitemsep} 
\usepackage{amsmath}   
\DeclareCaptionLabelSeparator{bar}{ \textbar\ }

\title{\textbf{Data-driven Prediction of Species-Specific Plant Responses to Spectral-Shifting Films from Leaf Phenotypic and Photosynthetic Traits}
\vspace{-0.5ex}}
\author{
\begin{minipage}{\textwidth}
\hspace*{0.5cm}  
\centering
\setstretch{1.2}
Jun Hyeun Kang$^{1}$, Jung Eek Son$^{2}$, Tae In Ahn$^{2,3}$\footnotemark[1]\\[0.5ex]
{\small
$^{1}$Smart Farm Research Center, Korea Institute of Science and Technology, Gangneung 25451, Republic of Korea\\[0.2ex]
$^{2}$Research Institute of Agriculture and Life Sciences, Seoul National University, Seoul 08826, Republic of Korea\\[0.2ex]
$^{3}$Department of Agriculture, Forestry and Bioresources, Seoul National University, Seoul 08826, Republic of Korea\\[0.2ex]
}
\end{minipage}
}

\date{}

\begin{document}
\raggedbottom

\maketitle

\renewcommand{\thefootnote}{\fnsymbol{footnote}}
\footnotetext[1]{Corresponding authors: tiahn@snu.ac.kr}

\begin{abstract}
The application of spectral-shifting films in greenhouses to shift green light to red light has shown variable growth responses across crop species. However, the yield enhancement of crops under altered light quality is related to the collective effects of the specific biophysical characteristics of each species. Considering only one attribute of a crop has limitations in understanding the relationship between sunlight quality adjustments and crop growth performance. Therefore, this study aims to comprehensively link multiple plant phenotypic traits and daily light integral considering the physiological responses of crops to their growth outcomes under SF using artificial intelligence. Between 2021 and 2024, various leafy, fruiting, and root crops were grown in greenhouses covered with either PEF or SF, and leaf reflectance, leaf mass per area, chlorophyll content, daily light integral, and light saturation point were measured from the plants cultivated in each condition. 210 data points were collected, but there was insufficient data to train deep learning models, so a variational autoencoder was used for data augmentation. Most crop yields showed an average increase of 22.5\% under SF. These data were used to train several models, including logistic regression, decision tree, random forest, XGBoost, and feedforward neural network (FFNN), aiming to binary classify whether there was a significant effect on yield with SF application. The FFNN achieved a high classification accuracy of 91.4\% on a test dataset that was not used for training. This study provide insight into the complex interactions between leaf phenotypic and photosynthetic traits, environmental conditions, and solar spectral components by improving the ability to predict solar spectral shift effects using SF.
\end{abstract}

\begin{figure}
    \centering
    \includegraphics[width=1\linewidth]{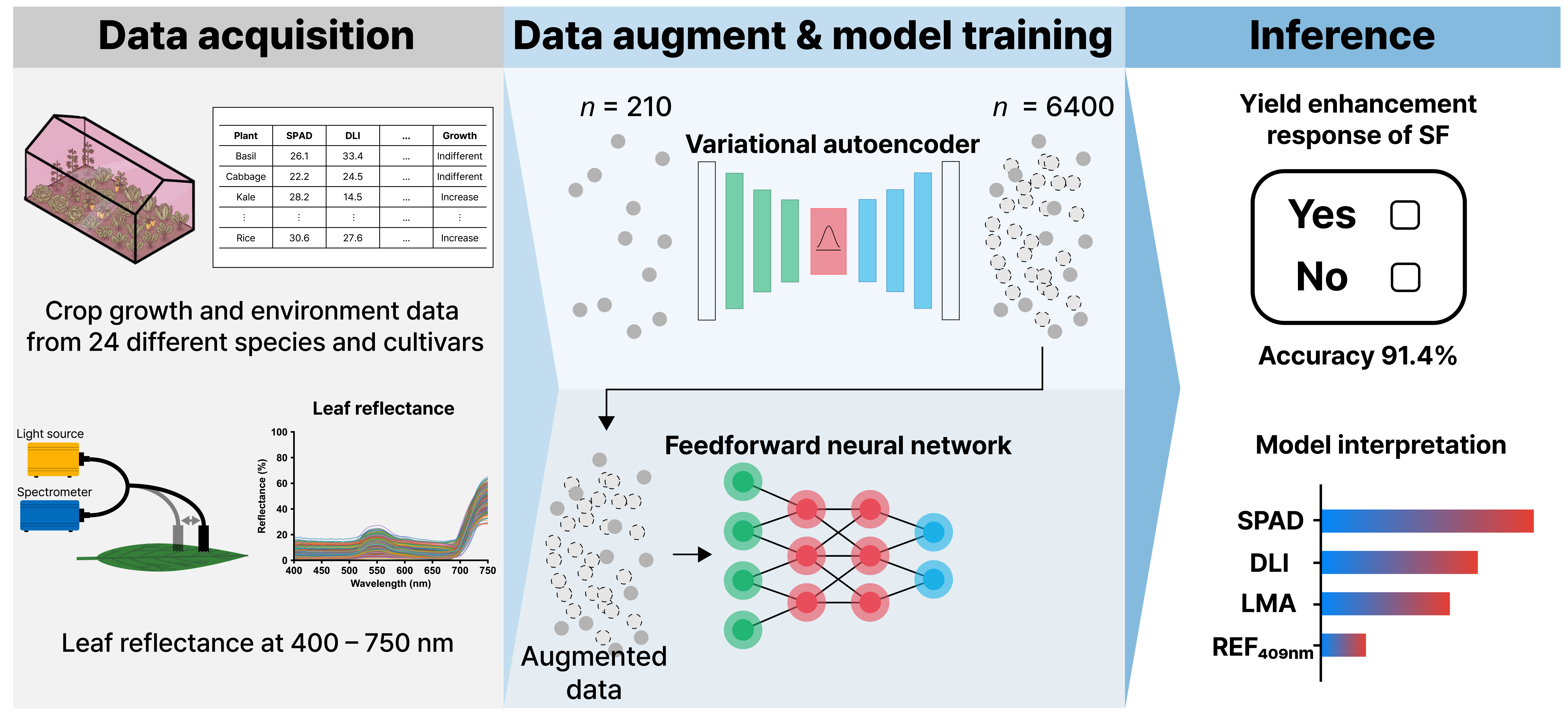}
    \caption{Graphical abstract of this study. The workflow is divided into three main steps. (1) Data acquisition: crop growth, environmental data, and leaf reflectance spectra (400–750 nm) were collected from 24 different species and cultivars grown in a greenhouse covered with a spectral-shifting (SF) film. (2) Data augment \& model training: to overcome the small dataset size (\textit{n}=210), a variational autoencoder (VAE) was trained to generate a large augmented dataset (\textit{n}=6400). This combined dataset was then used to train a feedforward neural network. (3) Inference: the final model predicts the binary (Yes/No) yield enhancement response to SF with 91.4\% accuracy. Furthermore, model interpretation using SHAP identified significant features for the prediction.}
    \label{fig:placeholder}
\end{figure} 

\section{Introduction}
The challenge of feeding the world's projected population of nearly 10 billion people by 2050 in a sustainable way is one of the greatest challenges of the 21st century\cite{Milla2017}. To meet the growing demand for food production while minimizing environmental impact, advanced agricultural technologies must be developed and implemented. Controlled environment agriculture (CEA), such as greenhouses and vertical farms, has emerged as a potential solution for enhancing crop production intensively and sustainably\cite{Cowan2022}. However, to maximize the potential of CEA, a comprehensive understanding of the complex interplay between environmental factors and plant responses is crucial. Among various environmental factors, light plays a critical role in regulating plant growth and development, with light intensity, quality, and photoperiod acting as key determinants of crop productivity\cite{Li2009,Izzo2020,Liu2021}. Therefore, optimizing light manipulation strategies in CEA holds great promise for improving crop yields and quality while reducing resource consumption and environmental footprint.

In a greenhouse, plant growth can be manipulated by adjusting the light environment with artificial lighting supplementation and spectral-selective films. These methods were used to optimize the light spectrum for photosynthesis and regulate photomorphogenesis\cite{Kim2023,Kang2022,Kwon2023}. However, artificial lighting supplementation may not be sustainable due to the additional energy input and CO$_2$ emissions. Moreover, spectral-selective films can reduce the intensity of specific wavelengths, leading to a decrease in overall light intensity. Spectral-shifting film (SF) uses fluorescent dyes in the film to shift green light into red light\cite{Paskhin2022}. Unlike spectrally selective films, SF converts light energy rather than blocking it, so less light is reduced. As a result, plants grown in sunlight with more red light due to SF exhibit an increased photosynthetic capacity and yield\cite{Li2017b, Yoon2020, Shen2021, Gao2022,Kang2022}. For instance, Chinese cabbage, lettuce, strawberries, and sweet pepper have shown an increase in plant yields\cite{Kang2022, Kang2023a, Gao2022, Kang2025}. However, the yields of some crops grown in SF were significantly reduced, and even within the same crop, the effect varied by cultivar\cite{Shen2021, Kang2023a, Yoon2021}. Thus, SF-induced improvement was species-specific and cultivar-specific.

Although SF could be an alternative for improving crop production, the inconsistent effects of SF on plant growth and yield are the main obstacle to the feasibility of solar spectrum modification. Considering one attribute of a crop has limitations in understanding the relationship between modified solar spectrum by SF and crop growth response. A previous study found that the photosynthetic capacity and productivity of Chinese cabbage increased when grown in the same polyethylene film (PEF) and SF, while lettuce did not show the same improvement\cite{Kang2023a}. The study identified chlorophyll components and leaf thickness as the factors responsible for this species specificity. Additionally, the photosynthetic enhancement effect of SF has been reported to increase the photosynthetic rate at light intensities above the light saturation point\cite{Yoon2020, Kang2022, Kang2023a, Kang2025}. Therefore, even if photosynthetic capacity is increased, if the light intensity during the cultivation period is low, this enhancement may be neutralized, potentially resulting in no growth enhancement effect from SF. Thus, relationships between various plant phenotypic traits, light environmental conditions, and SF-induced growth responses need to be elucidated. Machine learning and deep learning are widely used to understand complex causal relationships for a variety of phenomena\cite{Apolo-Apolo2020, Shin2022, Yoon2024}. These approaches have the potential to identify key plant characteristics that contribute to the species- and cultivar-specific effects of SFs, and to develop predictive models for optimizing SF application in diverse crop production systems.

To fill this research gap, the present study collected growth, phenotypic data, daily light integral, and photosynthetic parameters from 13 different crops and diverse cultivars between 2021 and 2024. Approximately 210 data points were collected over a period of four years. However, it is possible that there was insufficient data to train machine learning and deep learning models. Thus, the dataset was augmented using VAE, a type of generative artificial neural network model that trains the distribution of the data and generates new data. The average increase rate of crops in yield was about 22.5\% under SF compared to PEF. Various machine learning and deep learning models were created using the data obtained during the experiments to create a binary classifier that predicts the effect of SF. As a result, the feedforward neural network with augmented data achieved a high accuracy of 91.4\% on the test dataset that was not used for model training. This result enables the prediction of yield variances caused by SF application in relation to plant phenotypic traits, environmental conditions, and photosynthetic traits. This prediction can be used to determine the effectiveness of SF without the need for additional trial and error.

\section{Methods}

\subsection{Film properties}
Greenhouses were covered with either a conventional polyethylene film (PEF, Ultra film, Taekwangnewtech Co. Ltd., Seoul, Korea) or a spectral-shifting film (SF; Taekwangnewtech Co.). The relative transmittance spectra of PEF and SF were measured using a spectroradiometer (BLUE-Wave, StellarNet Inc., Tampa, FL, USA) (Fig. 2). The maximum absorption and emission wavelengths of SF were 576 nm and 622 nm, respectively. SF increased the transmittance of red light by 12.8\% compared to PEF. Consequently, the average transmittance in the photosynthetically active radiation (PAR) range for PEF and SF was 94.2\% and 90.8\%, respectively. 

\begin{figure}
    \centering
    \includegraphics[width=0.5\linewidth]{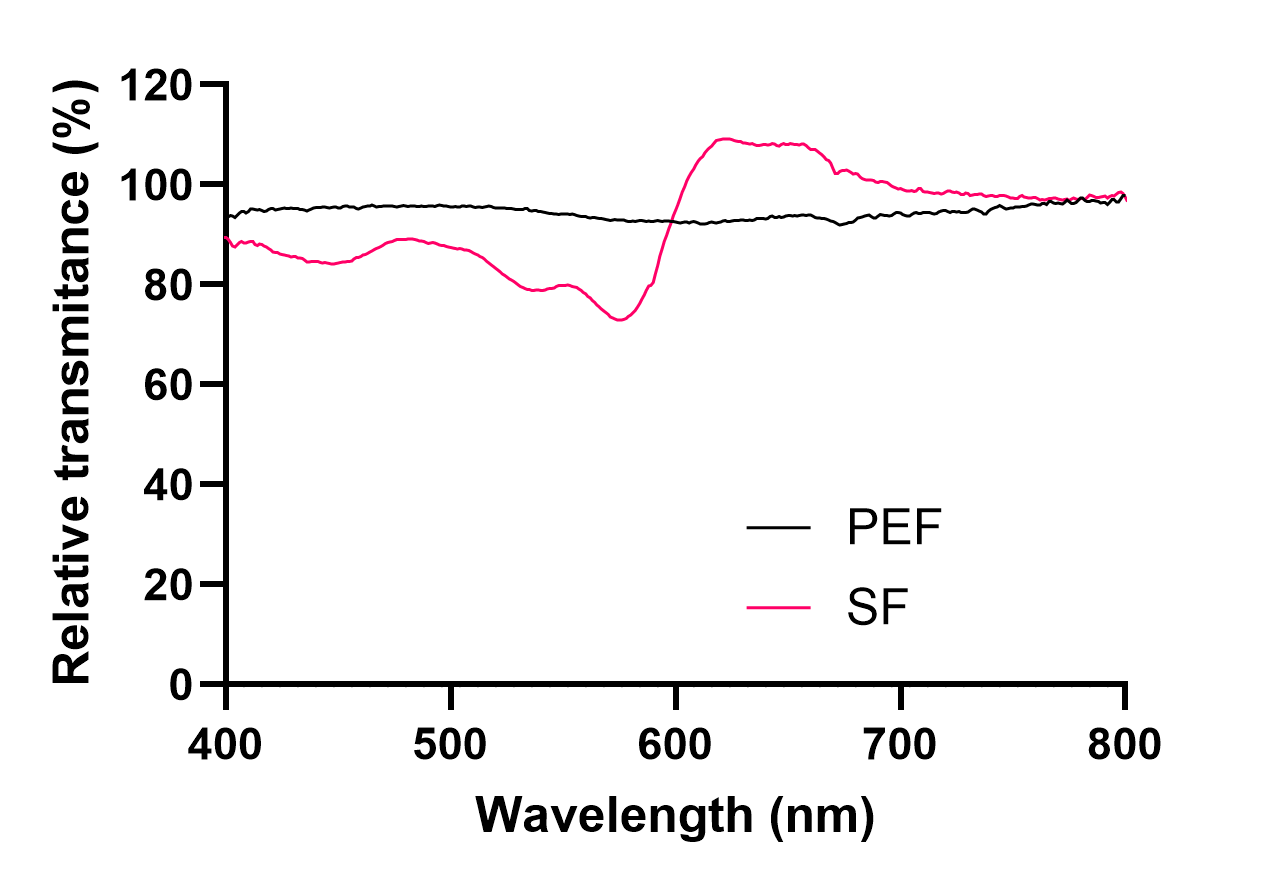}
    \caption{Comparison of relative transmittance between polyethylene film (PEF) and spectral-shifting film (SF).}
    \label{fig:placeholder}
\end{figure}

\subsection{Strawberry cultivation}
Sulhyang cultivar of strawberry (\textit{Fragaria} × \textit{ananassa} Duch.) was grown in commercial greenhouses situated in Miryang, Korea (35°36'N, 128°77'E). The greenhouses, measuring 95 m in length, 8.3 m in width, and 2.5 m in height, were covered with a double layer of film. The outer layer of all greenhouses consisted of PEF, while the inner layer was either PEF or SF. The study was conducted over three separate experiments, with new covering materials installed for each experiment: Experiment (Exp.) 1 from 18 Oct. 2021 to 4 Apr. 2022, Exp. 2 from 20 Oct. 2022 to 10 Apr. 2023, and Exp. 3 from 20 Oct. 2023 to 30 Apr. 2024.

Following the transplantation of strawberry seedlings into greenhouses containing coir growing medium with a 1:1 ratio of chip to dust, a nutrient solution specifically formulated for strawberries was administered four times daily, with an electrical conductivity (EC) ranging from 0.8 to 1.5 dS·$m^{-1}$. Each plant received 400 mL of the nutrient solution per day. The temperature during the day was regulated by adjusting the side windows of the greenhouse, while the temperature at night was maintained at a minimum range of 8 to 10℃ using an electric heater. To improve fruit quality and quantity, flower thinning was performed consistently across all treatments. The first flower cluster was thinned to retain 7 to 10 flowers per cluster, and subsequent clusters were thinned to maintain 5 flowers each. Leaf management was also implemented, with strawberry plants having 6 to 7 leaves at the onset of flowering in the first cluster and 7 to 8 leaves during the first fruit harvest. Additionally, old leaves were removed at regular intervals.

In Exp. 1, the cumulative strawberry yields were 2,525 and 2,723 kg in the two PEF greenhouses and 2,823 and 2,976 kg in the two SF greenhouses from 1 Nov. 2021 to 13 Apr. 2022. During Exp. 2, data on cumulative yields were collected from 9 Feb. to 10 Mar. 2023, with the two PEF greenhouses yielding 584 and 985 kg, while the four SF greenhouses produced 793, 834, 1,001, and 1,047 kg. For Exp. 3, the yield from 1 Jan. to 26 Mar. 2024 was 3,471 kg for the three PEF greenhouses combined and 4,176 kg for the three SF greenhouses combined.

\subsection{Pepper and watermelon cultivation}
Pepper (\textit{Capsicum annuum} L. cv. Cheongyang) of Exp. 1 was cultivated from 1 Oct. 2022 to 23 Dec. 2022 at the commercial greenhouse located in Miryang (35°36'N, 128°75'E), Korea. The outer layer of both greenhouses was covered with the PEF. The inner layer of one greenhouse was covered with PEF and that of the other greenhouse was covered with SF. Each greenhouse was 100 m in length and 4.4 m in width. The pepper seedlings were transplanted in soil mulched with black plastic film. Irrigation was provided through a drip irrigation system, and the plants were trained to grow in a V-shape. Peppers grown under SF exhibited a significantly higher leaf dry weight (56.9 ± 17.9 g) compared to those grown under PEF (38.4 ± 5.7 g). This represents a statistically significant increase of 48\% in leaf dry matter production. The cumulative yield of peppers grown in four PEF and four SF greenhouses was recorded at 6,907 and 8,067 kg, respectively. 

Following the harvest of peppers, the same greenhouses were used to cultivate watermelons (\textit{Citrullus lanatus} cv. Good Choice). The soil was mulched with plastic film before transplanting the watermelon seedlings. The seedlings were transplanted at 40 cm intervals. Drip irrigation tubing was installed underneath the plastic mulch to provide water and nutrients to the watermelon plants. The watermelon plants were managed using a two-stem training system, consisting of a main stem and a lateral stem. The main stem and lateral stem were trained in a T-shape, with the fruit set occurring on the main stem. For the watermelons, only the third female flower on the 18th to 21st node was retained, while all other flowers were removed, leaving only one fruit per plant. Four watermelons were randomly harvested from each of two greenhouses, one of which was a PEF greenhouse and the other a SF greenhouse. The fresh weights of the harvested watermelons were determined as yield. The average weight of watermelons grown in PEF was 6.4 ± 0.4 kg and that of watermelons grown in SF was 8.0 ± 0.7 kg.

Exp. 2 was initiated on 3 Aug. 2023 by transplanting pepper. The PEF film and SF were replaced with new ones on 2 Oct. Pepper yields were recorded from 6 Oct. to 21 Dec. 2023, and the cumulative yields were compared. The PEF greenhouses yielded 4,944 kg, while the SF greenhouses yielded 6,766 kg.

\subsection{Rice cultivation}
Rice (\textit{Oryza sativa} L. cv. Akibare) was cultivated from 1 June 2023 to 27 Oct. 2023 in a 3,960 m$^2$ paddy in Paju (37°82'N, 126°68'E), Korea. The rice seedlings were transplanted into the paddy with a spacing of 30 cm between rows and 15 cm between plants within each row. The field was irrigated regularly to maintain a water level of approximately 5–10 cm throughout the growing season. Fertilizers were applied according to the standard recommendations for rice cultivation. The small-scale greenhouse (2.0 m by 3.0 m) was installed on 12 June and covered with SF. For rice, the control of the experiment was set to the sample grown without cover. Ten plants from the control and SF treatment were randomly selected and measured the growth of rice plant. The harvested grain was oven dried and the dry weight of grain per plant was determined as the yield of rice. The dry weight of grain grown in control was 1.9 ± 0.3 g and that of grain grown in SF was 2.4 ± 0.6 g.

\subsection{Leafy rapeseed and spinach cultivation}
Rapeseed (\textit{Brassica napus} L.) was cultivated from 23 Nov. 2022 to 20 Feb. 2023 at the commercial greenhouse located in Chungju (37°07'N, 127°85'E), Korea. The two greenhouses were 100 m in length and 4.6 m in width and were covered with PEF and SF, respectively. The seed of rapeseed was sown in soil at each treatment without any additional heating. Irrigation was conducted using overhead sprinklers installed in the greenhouses. Ten plants from each greenhouse were randomly selected for harvesting. The leaf dry weight of rapeseed grown in PEF was 1.7 ± 0.4 g and that of rapeseed grown in SF was 2.8 ± 0.7 g. In the same greenhouse, spinach (\textit{Spinacia oleracea} L. cv. Winterstar) seeds were sown under the replaced PEF and SF on 5 Nov. 2023, and ten plants per greenhouse were harvested and examined on 10 Feb. 2024. Harvested plants were oven-dried, and leaf dry weight was quantified as the yield. The leaf dry weight of spinach grown in PEF was 2.9 ± 1.4 g and that of spinach grown in SF was 4.3 ± 1.4 g.

\subsection{Leafy vegetables and sweet pepper cultivation}
In a Venlo-type greenhouse of Seoul National University, located in Suwon (37°27'N, 126°99'E), Korea, experiments were conducted within small-scale greenhouse (2.6 m by 1.1 m) covered with PEF and SF. The greenhouse was maintained at a temperature of 24℃ during the day and 20℃ at night. The leafy vegetables were grown using hydroponic methods with a Hoagland nutrient solution maintained at the EC of 1.2 dS·$m^{-1}$. A diverse selection of leafy vegetables was cultivated, including Chinese cabbage (\textit{Brassica rapa} L. cv. ACC202 and Jinhongssam), pak choi (\textit{Brassica rapa} L. cv. Singsinghagye and Epikhigh), kale (\textit{Brassica oleracea} L. var. acephala cv. Manchoo collard), green and red lettuce (\textit{Lactuca sativa} L. cv. Butterhead, Caesars green, Topgreen, Jeock, Jeockchima, Oak, Superseonpung, Multihead), and basil (\textit{Ocimum basilicum} L. cv. Sweetbasil) were grown from 12 Nov. 2021 to 10 Feb. 2024. The sample size for each cultivar was organized as follows: five plants each for ACC202, Epikhigh, Jinhongssam, Manchoo collard, Singsinghagye, and Jeockchima; six plants each for Jeok and Oak; eight plants each for Butterhead, Yeolpung, Multihead, Sweetbasil, and Yeolpung; and ten plants each for Superseonpung and Topgreen.

The effect of SF on sweet pepper (\textit{Capsicum annuum} L. cv. Mavera) seedling was further investigated. The sweet pepper seeds were sown in rockwool trays with 240 plugs (Grodan Rockwool BV, Roermond, the Netherlands) on 3 Jan. 2023. After germination, the seedlings were covered with PEF and SF and then transplanted to rockwool blocks (Grodan PlanTop, Grodan Rockwool BV) on 2 Feb. 2023. The nutrient solution, with the EC ranging from 2.0 to 3.0 dS·$m^{-1}$, was provided to the plants to ensure optimal growth and development. The nursery was completed on 21 Feb. 2023, and ten plants per treatment were examined. The yield was determined by the shoot dry weight of sweet pepper seedlings. The shoot dry weight of sweet pepper seedling grown in PEF was 0.5 ± 0.1 g and that of sweet pepper seedling grown in SF was 0.8 ± 0.1 g.

To investigate the effect of modified solar spectrum on sweet pepper, the seedlings were transplanted in the Venlo-type greenhouse at Seoul National University on 22 Sep. 2022. PEF and SF were installed on the top and sides of the Venlo-type greenhouse, respectively, and an impermeable plastic sheet was installed in the center of the greenhouse to prevent the light interference. During the cultivation period, the greenhouse temperature was maintained between 20–26℃ using roof vents and hot-water pipe systems, and the relative humidity was controlled to be between 40–80\% using humidifiers. The nutrient solution concentration was 3.0 dS·m-1 for the vegetative stage and 4.0 dS·m-1 for the reproductive stage. Following bud formation, every plant was separated into two primary stems, which were then vertically trained along the V shape. Lateral stems were regularly pruned throughout the growing season. To promote early vegetative growth, the first and second flowers were eliminated across all treatments. Cultivation was completed on 20 Mar. 2023. During the cultivation period, the harvested fruits per plant were oven-dried and recorded. The yield of sweet pepper was determined as the dry weight of fruit per plant (\textit{n} = 37).

\subsection{Chlorophyll content and leaf reflectance analysis}
SPAD values were determined by measuring five randomly selected points on the upper surface of each plant, from the second or third leaf, using a SPAD meter and averaging the results (SPAD-502, Konica Minolta Inc., Tokyo, Japan). The chlorophyll (chl) a and b and carotenoid content were determined using three 2 mm leaf discs from the same leaf as the SPAD-measured leaf. The reflectance of each leaf was measured using the same leaf as the SPAD and chlorophyll analysis. The reflectance of leaf in the range of 400–750 nm was measured using the spectroradiometer (Blue-Wave spectrometer, StellarNet Inc., Tampa, FL, USA) connected with reflectance probe (R600, StellarNet Inc.) and tungsten halogen lamp (SL1, StellarNet Inc.). Spectral data typically requires preprocessing\cite{Yoon2023}. In this study, derivatives were used empirically to achieve the best performance.

\subsection{Daily light integral}
The daily integral was calculated based on the solar radiation data measured at regional weather stations during the cultivation period for each crop, considering the transmittance of the greenhouse film. The solar radiation data used in the calculations were obtained from the publicly available dataset provided by the Korea Meteorological Administration. 

\subsection{Light saturation point}
To measure the photosynthetic light response curve, a
portable photosynthesis system (LI-6400XT, Li-Cor, Lincoln, NE, USA) with an RB
LED light source (6400-02B, Li-Cor) was utilized. The leaves were first exposed
to a light intensity of 1,000 $\mu$mol$\cdot$m$^{-2}\cdot$s$^{-1}$ for 15 minutes
to allow for light acclimation. During the photosynthesis measurements, the
chamber conditions were maintained at a constant level, with a leaf temperature
of 25~$^{\circ}$C, 60\% relative humidity, and a CO$_2$ concentration of 400 ppm.
The light response curve was generated by measuring photosynthesis at various
light intensities, including 2,000, 1,500, 1,200, 900, 600, 400, 200, 100, 50,
and 0 $\mu$mol$\cdot$m$^{-2}\cdot$s$^{-1}$, in descending order. The measured
values were fitted to hyperbolic tangent photosynthesis model\cite{jassby1976} (Platt and
Jassby, 1976):

\begin{equation}
    P_n = P_{max} \times \tanh\left( \frac{I}{I_{sat}} \right) - R_d
\end{equation}

where $P_n$
is the net photosynthetic rate ($\mu$mol$\cdot$m$^{-2}\cdot$s$^{-1}$), $P_{max}$
is the maximum photosynthetic rate, $I$ is the photosynthetic photon flux density
($\mu$mol$\cdot$m$^{-2}\cdot$s$^{-1}$), $I_{sat}$ is the light saturation
point, and $R_d$ is the dark respiration rate ($\mu$mol$\cdot$m$^{-2}\cdot$s$^{-1}$).
    
During the experimental period, the photosynthetic light response curves were directly
measured for Chinese cabbage (cv. ACC202), lettuce (cv. Top Green, Butterhead,
and Multi leaf), kale strawberries, and sweet peppers. However, for the other
crops, photosynthesis measurements were not conducted. Therefore, for these
remaining crops, light response curves were extracted from previous studies and
fitted to the hyperbolic tangent photosynthesis model to calculate the light
saturation points. For basil\cite{Beaman2009, Solis-Toapanta2019},
Chinese cabbage\cite{Taylor2020}, eggplant\cite{Di2021},
lettuce\cite{Tsoumalakou2022, Zhou2022, Kong2024},
pak choi\cite{Hou2021, Lin2024}, pepper\cite{Masabni2016, Qiu2019},
potato\cite{Li2017a}, rapeseed\cite{Taylor2020, Dellero2021, Hsiao2023}, rice\cite{Xu2019, Du2020},
spinach\cite{Ikkonen2022, Sun2023, Tsoumalakou2023},
and watermelon\cite{Lu2020}, the light saturation points were derived
from photosynthetic light response curves reported in the cited studies.
To investigate the relationship
between the light saturation point and the average light intensity during the
cultivation period, the light saturation point offset (LSPO) was calculated by
subtracting the average light intensity during the cultivation period from the
calculated light saturation point for each crop.

\subsection{Statistical analysis}
The statistical differences of yield grown at PEF and SF were analyzed using R software (R 3.6.2, R Foundation, Vienna, Austria). Prior to comparing the means, the homogeneity of variances for each variable was evaluated using an \textit{F}-test. Following the confirmation of homogeneity, a Student's \textit{t}-test was conducted to determine the significance of differences between the means of PEF or control and SF treatments. The threshold for statistical significance was set at \textit{P}-value$~<~$0.05.

\subsection{Development of machine learning and deep learning classifiers}
The efficacy of SF on plant yields was predicted using machine learning and deep learning models. To this end, a binary classification was set as the goal, whereby the \textit{P}-value was predicted to be less than 0.05 using a \textit{t}-test analysis for each crop. The input parameters for the machine learning and deep learning models included leaf phenotypic traits such as chlorophyll content, leaf reflectance, and SPAD values, as well as environmental factors including DLI and LSPO. The study employed decision tree-based models, including a basic decision tree, ensemble models such as random forest and XGBoost, and designed a deep learning model using a feedforward neural network (FFNN), in comparison to the fundamental logistic regression model. The experimental procedure is depicted in Fig. 3. The number of data is critical to train the FFNN. To address the lack of data, the experiment utilized data augmentation techniques, specifically using a variational autoencoder (VAE). VAE is a generative model that learns the distribution of the trained data and generates new data within the trained distribution range. The VAE architecture consisted of five layers in both the encoder and decoder. The number of neurons in each layer was set to 256, 128, 64, 32, and 16, respectively (Fig. 4). The latent dimension, which represents the compressed representation of the input data, was set to 7. The learning rate for VAE was 0.001. To mitigate overfitting, batch normalization and a dropout rate of 0.1 were applied to each layer of the VAE. The data augmentation process generated four augmented datasets, consisting of 400, 800, 1,600, and 6,400 samples, respectively. Each model was trained using the original data and the four augmented datasets. The trained models were then evaluated using the separate test set of crop data that was not included in the training process (Table 1). The hyperparameters of decision tree (DT), random forest (RF), XGBoost were fine-tuned using the BayesSearchCV library of scikit-optimize (Table 2). For each trained model, the feature importance was calculated using SHapley Additive exPlanations (SHAP). SHAP is a game-theoretic approach to explain the output of machine learning models by assigning importance scores to each feature\cite{Lundberg2017}. The SHAP values provide insight into the contribution of each feature to the model predictions, allowing for a better understanding of the factors that influence the effectiveness of SF on crop yield. Given the disparate levels of significance attributed to each model, MinMax normalization was employed to facilitate a comparative analysis of the models' relative importance.

\begin{figure}
    \centering
    \includegraphics[width=0.7\linewidth]{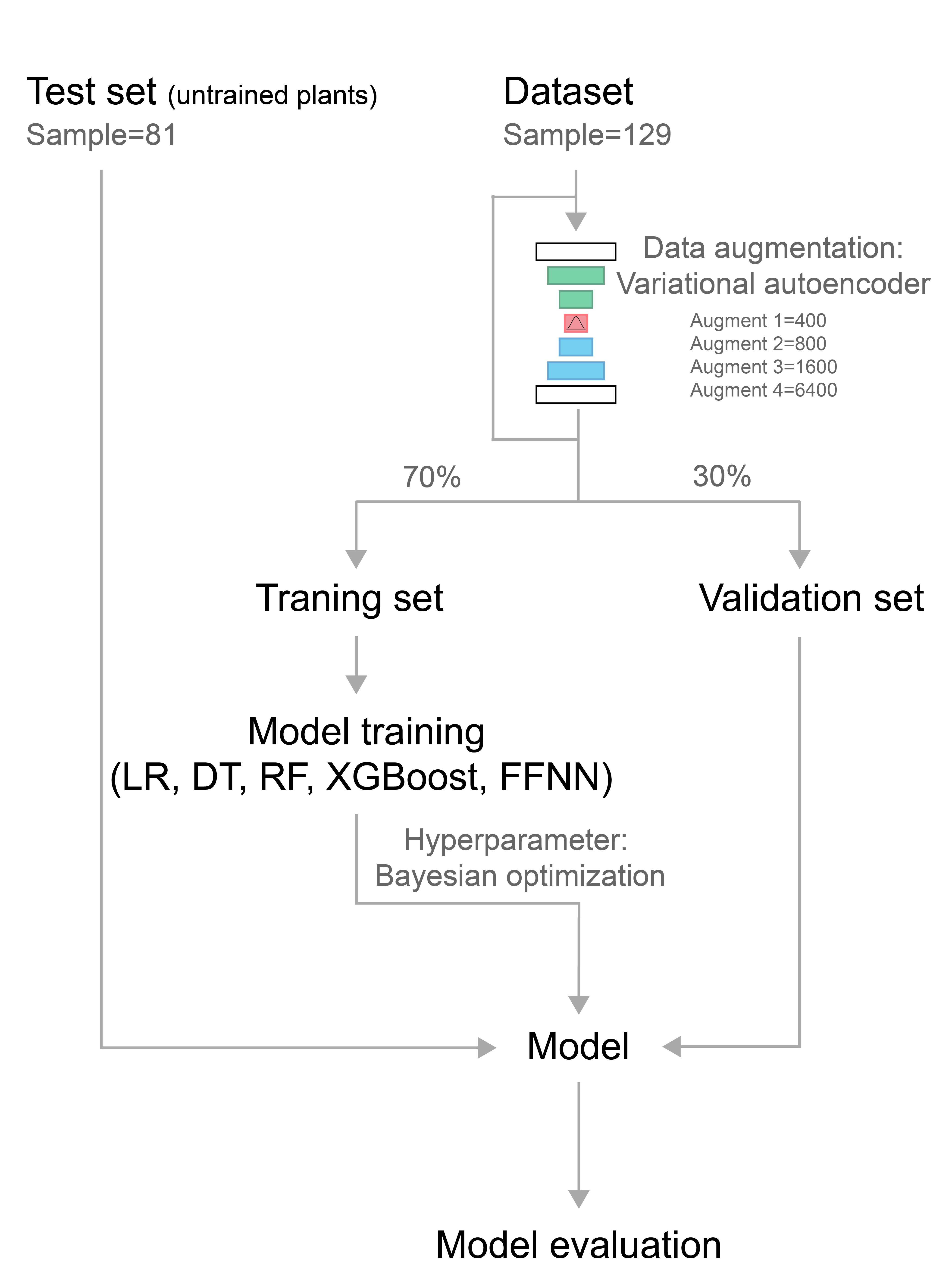}
    \caption{Schematic diagram of experimental procedure to develop the binary classification model of this study.}
    \label{fig:placeholder}
\end{figure}

\begin{figure}
    \centering
    \includegraphics[width=0.5\linewidth]{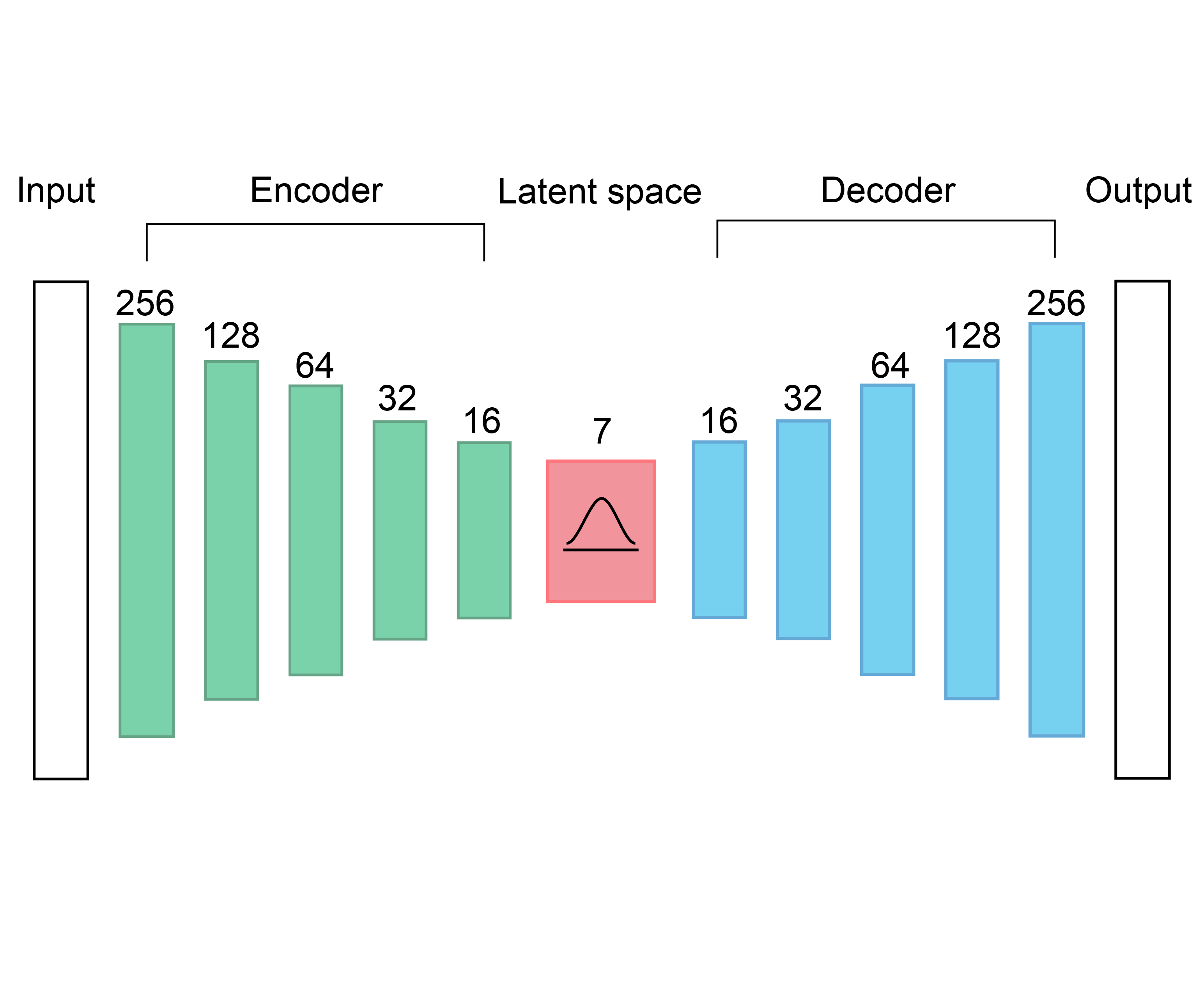}
    \caption{Visual representation of a variational autoencoder for data augment. }
    \label{fig:placeholder}
\end{figure}

\begin{table}[ht] 
\renewcommand{\arraystretch}{0.6} 
\centering
\caption{Plant and cultivars used in the train and test datasets.} 
\label{tab:dataset_list} 
\begin{tabular}{lll}
\toprule
\textbf{Dataset} & \textbf{Plant} & \textbf{Cultivar} \\
\midrule
\multirow{6}{*}{Train set} & Chinese cabbage & ACC202, Jinhongssam \\
                           & Kale              & Manchoo collard \\
                           & Leafy rapeseed    & Unknown \\
                           & Lettuce           & Caesars green, Jeockchima, Oak, Topgreen, Superseonpung \\
                           & Pak choi          & Epik high, Singsinghaye \\
                           & Strawberry        & Sulhyang \\
\midrule 
\multirow{9}{*}{Test set} & Pepper            & Cheongyang \\
                           & Sweet pepper      & Mavera \\
                           & Basil             & Sweet basil \\
                           & Eggplant          & Black pearl \\
                           & Lettuce           & Butterhead, Multihead, Yeolpung \\
                           & Potato            & Sumi \\
                           & Rice              & Akibare \\
                           & Spinach           & Winterstar \\
                           & Watermelon        & Good choice \\
\bottomrule
\end{tabular}
\end{table}

\begin{table}[ht]
\renewcommand{\arraystretch}{0.6} 
\centering
\caption{Search spaces of hyperparameters of machine learning and deep learning models.} 
\label{tab:hyperparams} 
\begin{tabular}{lll}
\toprule
\textbf{Model} & \textbf{Hyperparameter} & \textbf{Search space} \\
\midrule
\multirow{5}{*}{Decision tree} & Criterion & Gini or entropy \\
                               & Max depth & 3--6 \\
                               & Min samples split & 10--400 \\
                               & Min samples leaf & 10--400 \\
                               & Max feature & 10--50 \\
\midrule
\multirow{5}{*}{Random forest} & Number of estimators & 50--300 \\
                               & Max depth & 6--12 \\
                               & Min samples split & 30--400 \\
                               & Min samples leaf & 30--400 \\
                               & Max feature & 10--100 \\
\midrule
\multirow{7}{*}{XGBoost} & Number of estimators & 50--300 \\
                         & Max depth & 6--12 \\
                         & Min child weight & 5--200 \\
                         & Alpha & 1.5--400 \\
                         & Gamma & 2--400 \\
                         & lambda & 1.5--400 \\
                         & Colsample bytree & 0.1--0.5 \\
\midrule
\multirow{5}{*}{\shortstack{Feedforward neural \\ network}} & Number of layers & 1--2 \\
                               & Hidden dimension & 2--32 \\
                               & Weight decay & $1\text{e}^{-5}$--$1\text{e}^{-2}$ \\
                               & Learning rate & $1\text{e}^{-7}$--$1\text{e}^{-3}$ \\
                               & Batch size & 16--512 \\
\bottomrule
\end{tabular}
\end{table}

\subsection{Model evaluation}
The augmented data generated by VAE was visualized
using dimensionality reduction with principal component analysis (PCA) and
evaluated using Wasserstein distance to ensure that the augmented data
represented the original data well. The Wasserstein distance, which measures
the distance of the probability distribution of the original ($\mu$) and the
augmented data ($\nu$) from each other, was used to ensure that the
augmented data could be a representation of the original data.
\begin{equation}
    \text{Wasserstein distance } (\mu, \nu) = \left( \inf_{\gamma \in \Gamma(\mu, \nu)} \mathbb{E}_{(x,y) \sim \gamma} [d(x,y)^p] \right)^{1/p}
\end{equation}
Where $\Gamma(\mu, \nu)$ is the set of all couplings of $\mu$ and $\nu$. $\mathbb{E}$ represents the expected value with respect to the
joint probability distribution $\gamma$. $d(x,y)$ is a distance function that measures the distance
between points $x$ and $y$ in a given metric space. $p$ represents the order of the $L_p$
norm.

The prediction result of the trained model
can be classified into four metrics: 1) true positive (TP), where the model
correctly predicts the positive class; 2) true negative (TN), where the model
correctly predicts the negative class; 3) false positive (FP), where the model
incorrectly predicts the positive class (Type I error); and 4) false negative
(FN), where the model incorrectly predicts the negative class (Type II error).

To evaluate the trained models, two
primary metrics were employed: accuracy and the area under the receiver operating
characteristic curve (AUC).

Accuracy is the most widely used metric
for measuring the overall correctness of the model's predictions and is
calculated using the following formula:
\begin{equation}
    \text{Accuracy} = \frac{TN + TP}{TN + FN + TP + FP}
\end{equation}
On the other hand, AUC provides a
comprehensive assessment of the performance of model to distinguish between
positive and negative classes across various classification thresholds. It is
derived from the receiver operating characteristic (ROC) curve, which plots the
True Positive Rate (TPR) against the False Positive Rate (FPR) at different
classification thresholds. The formulas for TPR and FPR are:
\begin{align}
    \text{TPR} &= \frac{TP}{TP + FN} \\
    \text{FPR} &= \frac{FP}{FP + TN}
\end{align}
A higher AUC value indicates better model
performance, with an AUC of 1 representing a perfect classifier. By utilizing
both accuracy and AUC, the trained models can be thoroughly evaluated to
determine their effectiveness in predicting the impact of SF on plant yields.

\section{Results}
\subsection{SF-modified sunlight enhanced yield of crops, but not universal}
To investigate the response of different plants to
modified sunlight, empirical experiments was conducted on a variety of leafy
greens, fruits and vegetables, and grains. These experiments were conducted in
four regions of South Korea (Fig. 5). The SF-modified sunlight had a
variety of effects on the plants. The shoot dry weight of basil (sweet basil)
exhibited a 5.5\% reduction in response to modified sunlight, while green
Chinese cabbage (ACC202) showed a significant increase of 38.3\%. In contrast,
the shoot dry weight of red Chinese cabbage (Jinhongssam) exhibited a 1.1\%
reduction in comparison to PEF, and that of kale (Manchoo collard) exhibited a
12.7\% reduction. The leaf dry weight of leafy rapeseed grown with SF was found
to be significantly increased by 67.7\%. Among lettuce varieties, Jeock, Superseonpung,
and Topgreen showed significant increases of 19.8\%, 20.7\%, and 23.2\%,
respectively. However, Butterhead, Caesar Green, Jeockchima, Multihead, Oak,
and Yeolpung showed no difference in growth or even decreased. Both green and
red pak choi, Epik High and Singsinghagye, showed growth increases of 15.2\% and
3.4\%, respectively, but the difference was not significant. The leaf dry weight
of spinach (Winterstar) was significantly increased by 68.6\%.

\begin{figure}
    \centering
    \includegraphics[width=0.9\linewidth]{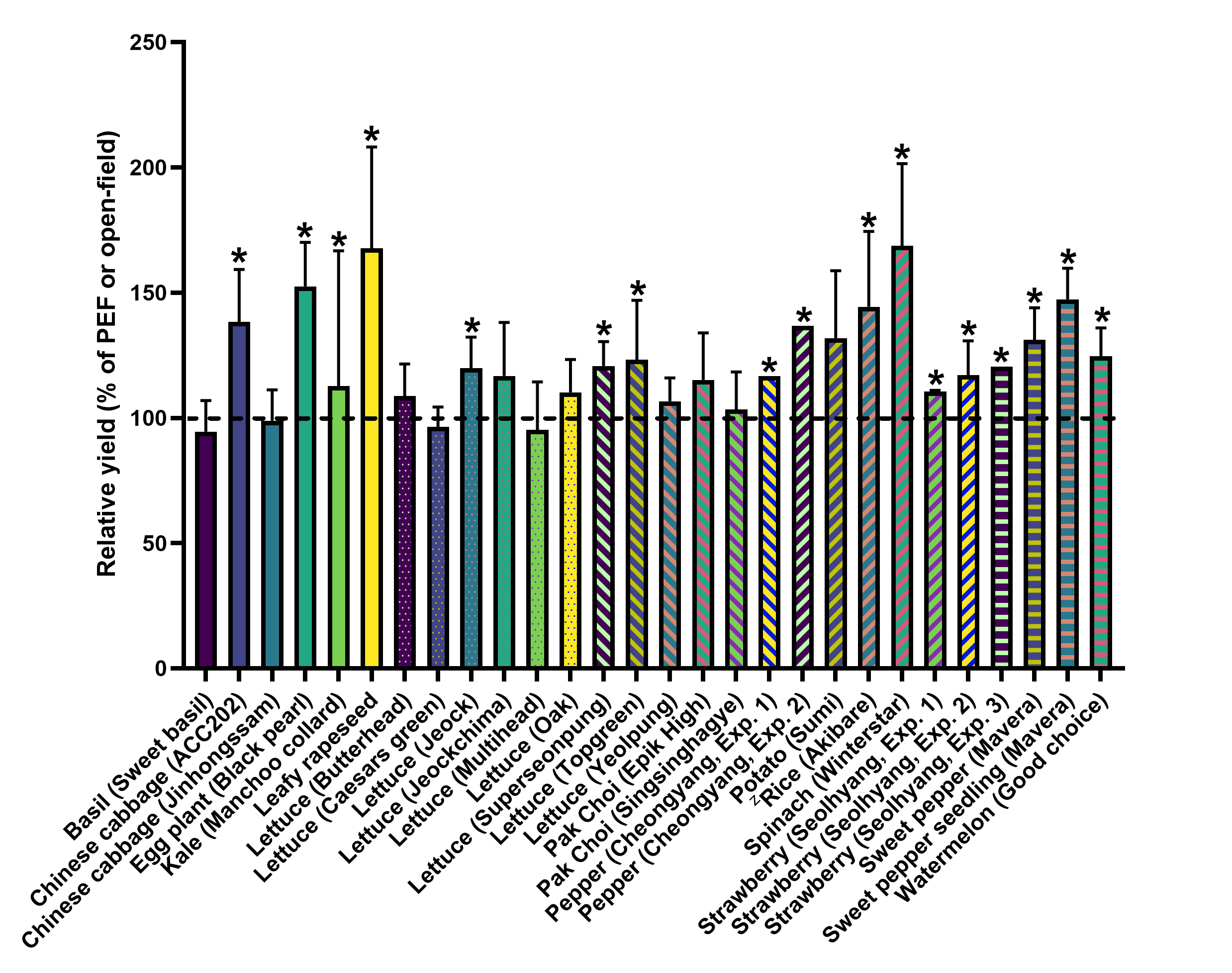}
    \caption{Relative
yield of crops grown at polyethylene film (PEF) or no cover versus spectral-shifting
film (SF). Bars without error bars are average. Data are Means$\pm$S.D.
The asterisks indicate significant differences (Student's \textit{t}-test, $^*$\textit{P}-value~$<$~0.05). 
\textsuperscript{z}Rice was set to open-field as a control.}
    \label{fig:placeholder}
\end{figure}

In the case of rice (Akibare), the dry weight of grain
grown under SF was significantly increased by 44.3\% compared to the control
group established without the use of any film. The shoot dry weight of
potato (Sumi) plant increased by 31.9\% under SF. In addition, the cumulative
fruit weight of pepper (Cheongyang) in Exps. 1 and 2 under SF increased by
16.8\% and 36.8\%, respectively, when grown under SF compared to those grown
under PEF. The shoot dry weight of eggplant (Black pearl) grown under SF
conditions was significantly increased by 52.5\%. The cumulative fruit yield of
strawberry (Sulhyang) grown under SF treatment increased by 10.5\%, 17.1\% and 20.3\%
in Exps. 1, 2, and 3, respectively, compared to those grown under PEF. The fruit
yield of sweet pepper (Mavera) grown under SF treatment significantly increased
by 36.8\% compared to those grown under PEF. The shoot dry weight of sweet
pepper (Mavera) seedlings grown under SF treatment increased significantly by
47.3\% compared to those grown under PEF. Finally, the fruit weight of
watermelon (Good choice) was significantly increased by 24.6\% under SF. Following
the empirical experiments of 24 crops, it was determined that the application
of SF resulted in an average increase in yield of 22.5\%. These findings suggest
that modifying the solar spectrum can enhance yields in a diverse range of
crops. The calculated light saturation points and LSPO for each crop are shown
in Table 3. Among all crops, lettuce had the lowest light saturation point,
while Chinese cabbage and rice had the highest light saturation points. The
calculated LSPO varied depending on the weather, climate, and season at the
time the crop was grown.

\begin{table}[H]
\centering
\renewcommand{\arraystretch}{0.6} 
\caption{The calculated light saturation point, light saturation point offset, and statistical significance of differences between control and spectral-shifting film for various plants. } 
\label{tab:lspo_data}
\begin{tabular}{lccc} 
\toprule
\textbf{Plant} & \textbf{\shortstack{Light saturation point}} & \textbf{LSPO\textsuperscript{z}} & \textbf{Significant\textsuperscript{y}} \\
\midrule
Basil & 468 $\pm$ 36\textsuperscript{x} & 303 & 0 \\
\midrule
Chinese cabbage & & & \\
 \hspace{0.5cm} cv. ACC202 & 725 & 539 & 1 \\
 \hspace{0.5cm} cv. Jinhongssam & 689 & 504 & 0 \\
\midrule
Kale & 470 & 14 & 1 \\
\midrule
Leafy rapeseed & 567 $\pm$ 76 & 379 & 1 \\
\midrule
Lettuce & & & \\
 \hspace{0.5cm} cv. Butterhead & 263 & 100 & 0 \\
 \hspace{0.5cm} cv. Caesars green & 251 & -205 & 0 \\
 \hspace{0.5cm} cv. Jeock & 230 $\pm$ 7 & -154 & 1 \\
 \hspace{0.5cm} cv. Jeockchima & 251 & -205 & 0 \\
 \hspace{0.5cm} cv. Multihead & 468 & 296 & 0 \\
 \hspace{0.5cm} cv. Oak & 230 $\pm$ 7 & -217 & 0 \\
 \hspace{0.5cm} cv. Superseonpung & 293 & -85 & 1 \\
 \hspace{0.5cm} cv. Topgreen & 220 & -158 & 1 \\
 \hspace{0.5cm} cv. Yeolpung & 251 & 188 & 0 \\
\midrule
Pak choi & & & \\
 \hspace{0.5cm} cv. Epik High & 490 $\pm$ 262 & 303 & 0 \\
 \hspace{0.5cm} cv. Singsinghaye & 490 $\pm$ 262 & 301 & 0 \\
\midrule
Spinach & 412 & 219 & 1 \\
\midrule
Rice & 713 $\pm$ 50 & 352 & 1 \\
\midrule
Potato & 359 $\pm$ 124 & 106 & 0 \\
\midrule
Pepper & 499 & 210 & 1 \\
\midrule
Eggplant & 389 $\pm$ 10 & 20 & 1 \\
\midrule
Strawberry & 432 & 157 & 1 \\
\midrule
Sweet pepper & 546 & 302 & 1 \\
\midrule
Watermelon & 478 & 154 & 1 \\
\bottomrule
\end{tabular}
\par 
\begin{flushleft} 
\footnotesize  
\textsuperscript{z}LSPO: light saturation point offset, calculated as light saturation point – average light intensity during each crop cultivation.  \\
\textsuperscript{y}Significance: 1 indicates significant difference (\textit{P}-value~$<$~0.05) between SF and control treatments based on t-test, 0 indicates no significant difference between treatments.  \\
\textsuperscript{x}Means $\pm$ S.D. 

\end{flushleft}
\end{table}

\subsection{Data augmentation with variational autoencoder}
A total of 210 data points were collected from experiments conducted with 24 different cultivars over 3 years. However, this number of data was not enough data to train machine learning and deep learning models. Thus, data augmentation with VAE was implemented (Fig. 6). The histograms of the distributions of the raw data and augmented data exhibited a high degree of similarity (Fig. 6a–g). Upon reduction of the high-dimensional characteristics of each sample to two dimensions using PCA and subsequent visualization in two dimensions, the similar distributions of the raw and augmented data were observed (Fig. 6h). The Wasserstein distance, which is a metric of the degree of similarity between the data generated by the generative model and the original data, was also found to be close to zero (Fig. 6i).

\begin{figure}
    \centering
    \includegraphics[width=1\linewidth]{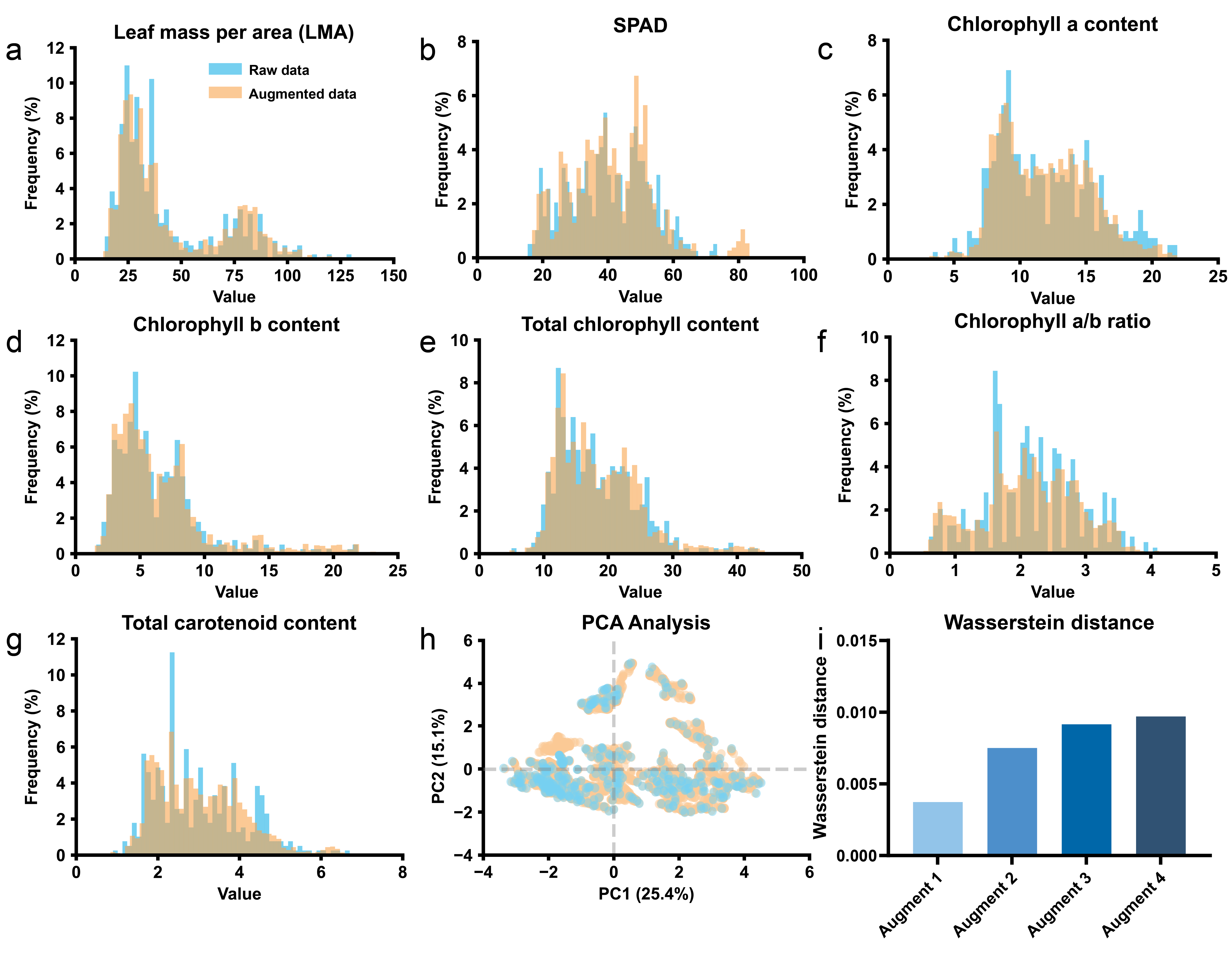}
    \caption{Histogram of the phenotypic data of leaves from the experiments (a–g). 2D visualization of high-dimensional leaf phenotypic data and augmented data using PCA (h). Wasserstein distance between original data and augmented datasets (i).}
    \label{fig:placeholder}
\end{figure}

\subsection{Model performance to predict the yield enhancement effect of SF}
LR trained on original and augmented data showed high validation accuracy. The highest test accuracy, consisting of plants not used in training, was 81.5\% on augment 4 data (Fig. 7). DT tended to have lower validation accuracy and lower test accuracy among machine learning models. The ensemble model, RF, had a higher test accuracy than DT, with 84.0\% on augment 2 and 4. XGBoost, another ensemble model, outperformed LR and DT, but had less predictive power than RF. XGBoost had a test accuracy of 82.7\% on augment 2. FFNN, the deep learning model, had higher validation accuracy than machine learning models. However, the test accuracy was low until augment 2 with less data, and then the FFNN model trained with the datasets from augment 3 and 4 predicted the test crops with high accuracy. Specifically, the FFNN model trained using augment 4 was able to predict the crops that would respond to SF with the accuracy of 91.4\% using leaf phenotypic and photosynthetic traits, and environmental conditions.

\begin{figure}
    \centering
    \includegraphics[width=1\linewidth]{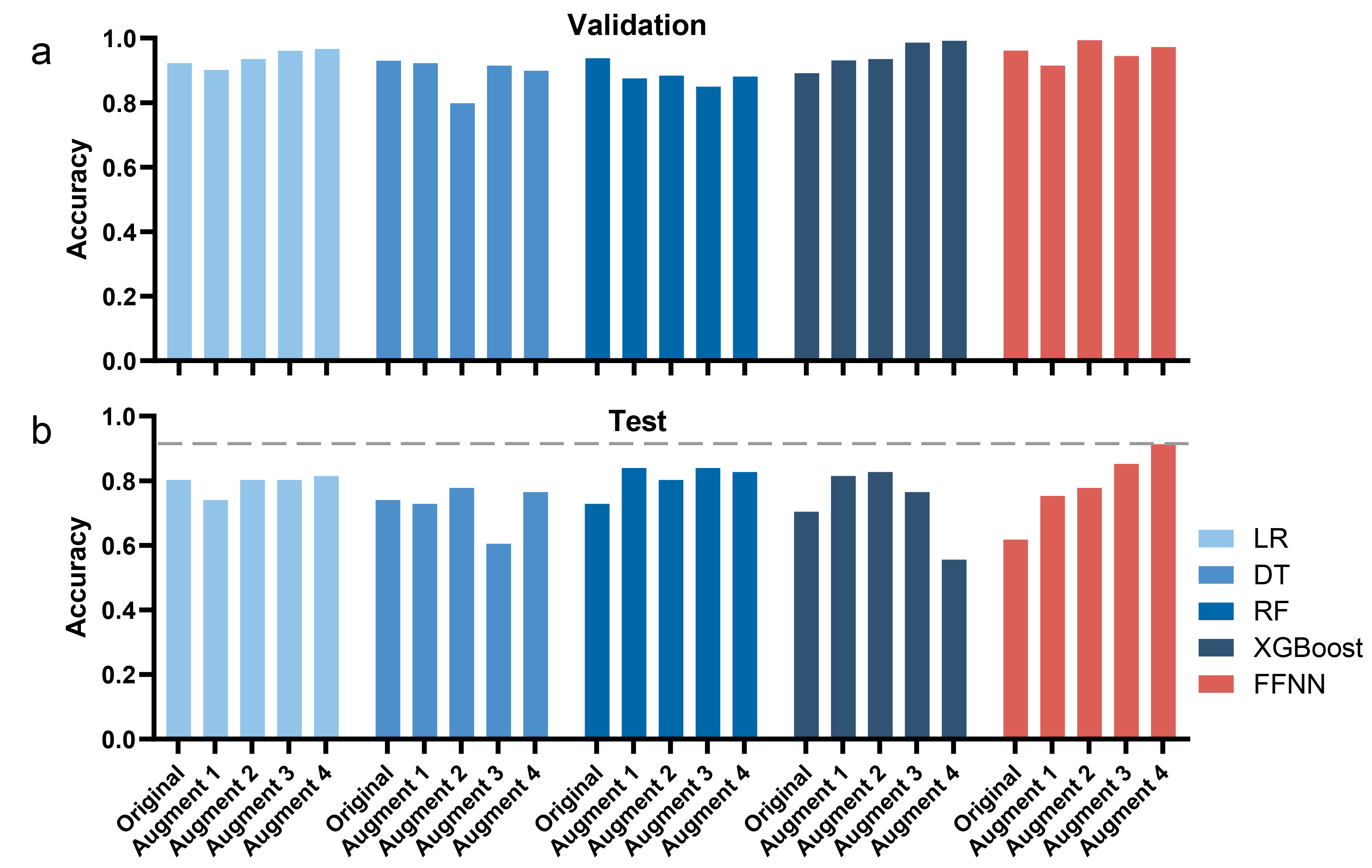}
    \caption{Classification performance of the logistic regression (LR), decision tree (DT), random forest (RF), XGBoost, and feedforward neural network (FFNN) trained using the original data and four augmented datasets.}
    \label{fig:placeholder}
\end{figure}

Fig. 8 shows the ROC curves for the best-performed model on each dataset, selected based on the highest test accuracy. Similar to the accuracy, the AUC of FFNN showed the highest score with 0.97. The LR, RF, XGBoost and DT models followed FFNN in performance, with their AUC values ranked in descending order, but did not show comparably strong results. Further investigation revealed that the FFNN trained through augment 4 could predict with high probability which plants would be effectively cultivated under SF (Fig. 9). Specifically, it correctly identified 31 out of 37 data that did not show growth enhancement effects under the SF, while misclassifying 6. Among the 44 data that showed positive effects from the SF, it correctly classified all but one case.

\begin{figure}
    \centering
    \includegraphics[width=0.5\linewidth]{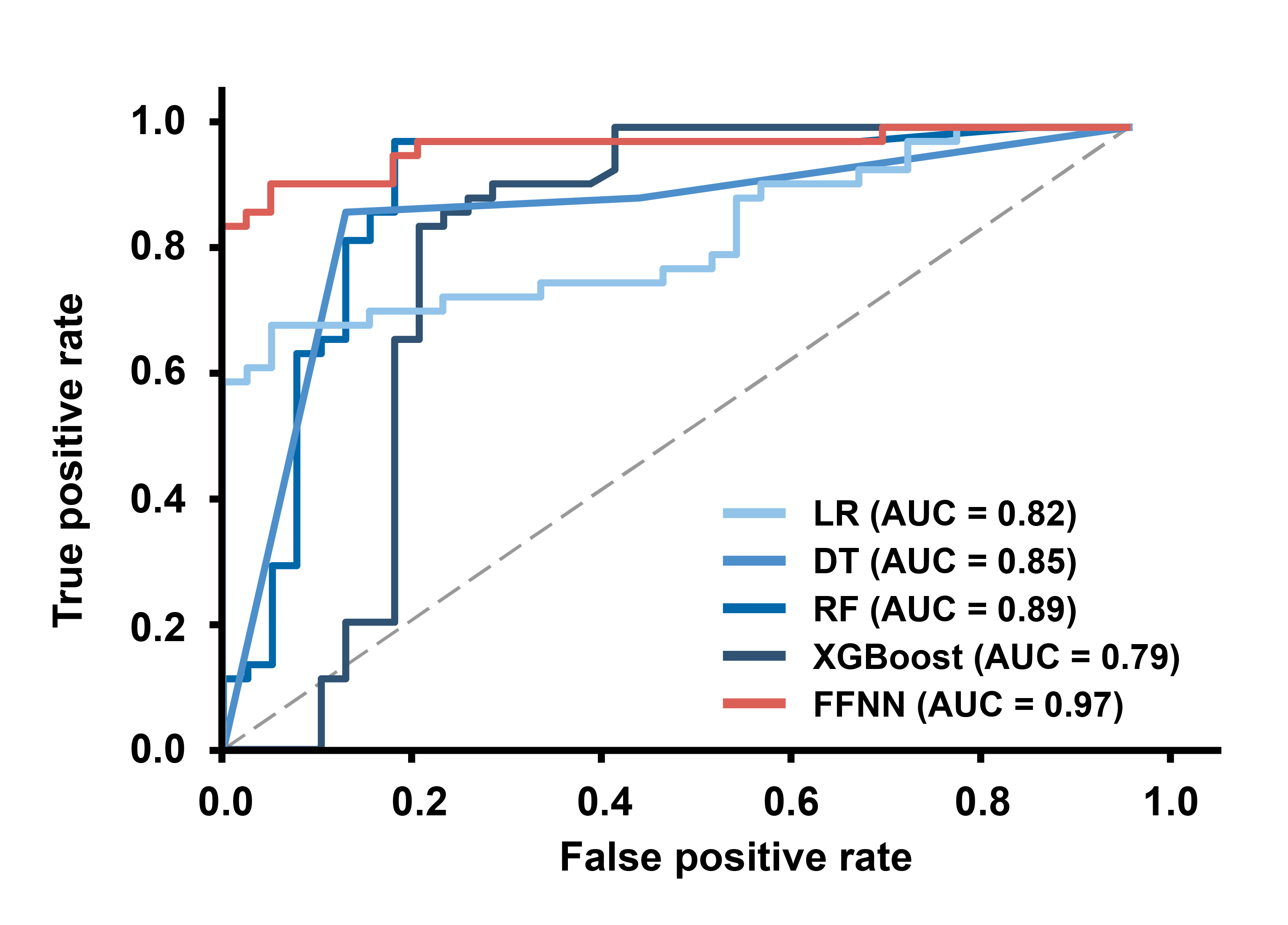}
    \caption{Receiver operating characteristic (ROC) curves and area under the ROC curve (AUC) of the logistic regression (LR), decision tree (DT), random forest (RF), XGBoost, and feedforward neural network (FFNN) trained using the original data and four augmented datasets.}
    \label{fig:placeholder}
\end{figure}

\begin{figure}
    \centering
    \includegraphics[width=0.5\linewidth]{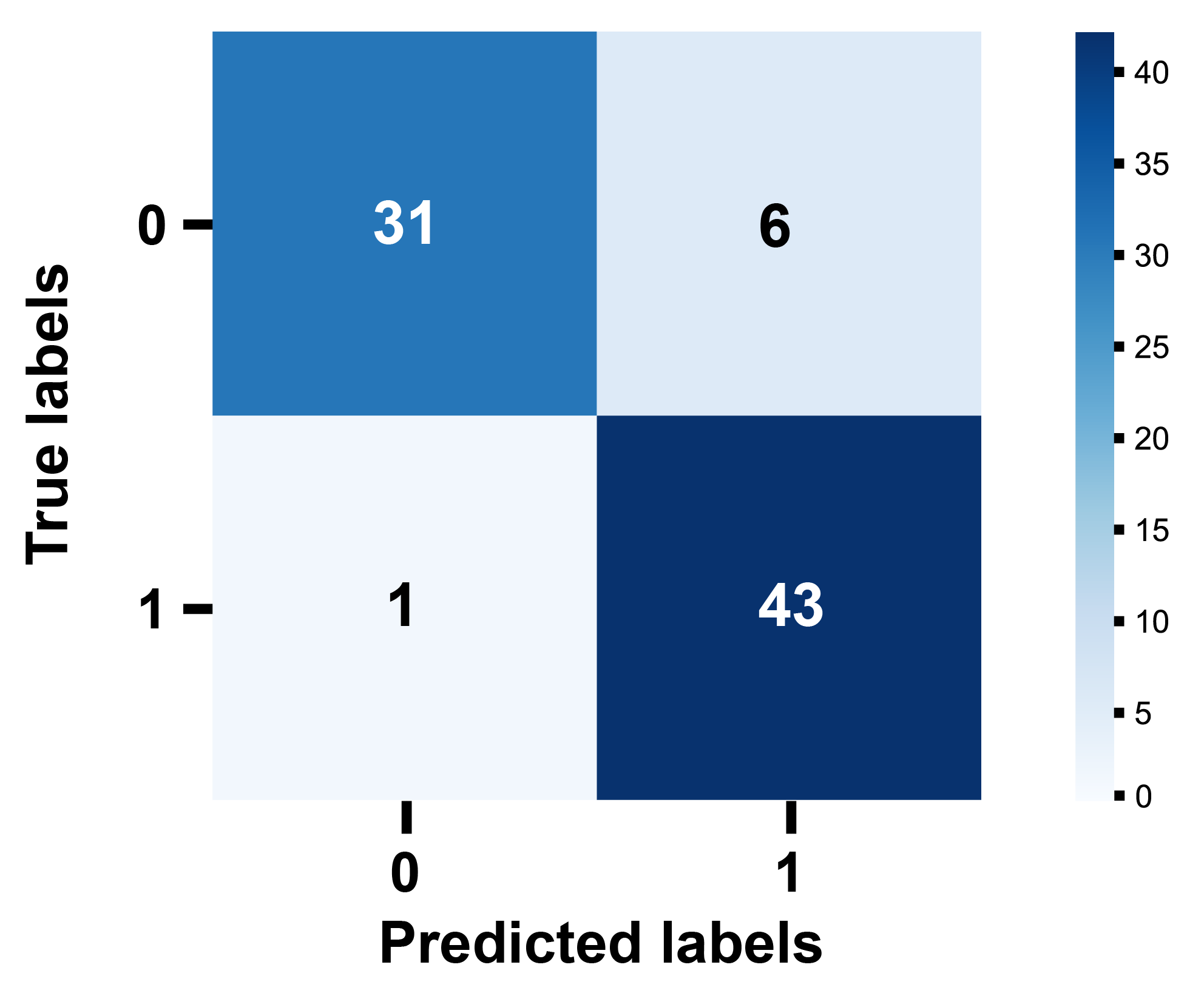}
    \caption{Confusion matrix of feedforward neural network (FFNN) trained using the augment 4 dataset.}
    \label{fig:placeholder}
\end{figure}

\subsection{Features that affect the inference of the models}
The SHAP values of the best performing models on each dataset, selected based on the highest test accuracy, are shown in Fig. 10. For the LR, it was determined that SPAD and LMA had the greatest impact on the predictive power of the model, with DLI and LSPO also being important. Regarding reflectance by wavelength, LR considered reflectance from 400 to 420 nm, 500 to 540 nm, and 690 to 750 nm to be of particular importance. For DT, only two reflectance, 700 and 742 nm, were used in the model's predictions. For RF, 693 nm has the highest importance, followed by 625 nm, and then the importance is distributed in the range of 691 nm to 701 nm. XGBoost considered the reflectance at 696 nm to be the most important feature. It then showed a high peak at 414, 626, and 695 nm to be of high importance. Chl b also showed high importance. In the best prediction model, which is FFNN, importance and directionality of the data are shown in Fig. 11. This plot shows the impact of each feature on the model’s prediction of SF effectiveness. The horizontal position of each point shows the impact on the model output, with points to the right of 0 indicating a positive influence on SF effectiveness prediction, and points to the left indicating a negative influence. The analysis reveals that SPAD, chl b, and total chl content are among the most influential features for predicting the effect of SF on growth. Other important features include the DLI, chl a/b ratio, and LMA. Additionally, the LSPO also seems to have a noticeable impact on the model's predictions, although it has a more complex relationship, as evidenced by the mixed distribution of red and blue points. In terms of reflectance, the lower reflectance at from 694 to 703 nm and the higher reflectance at 408 and 409 nm, the model predicted an effect of SF.

\begin{figure}
    \centering
    \includegraphics[width=0.8\linewidth]{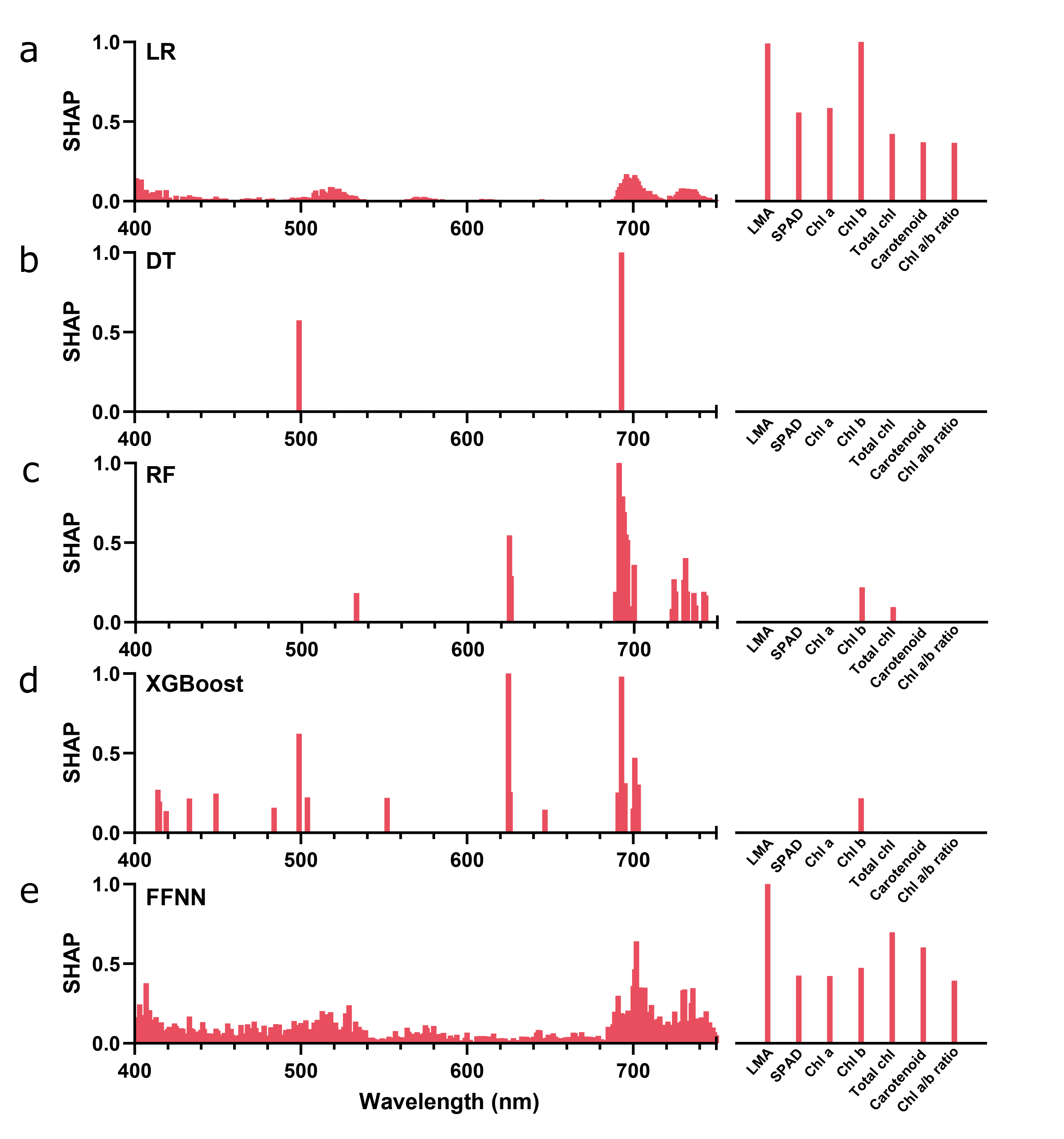}
    \caption{Normalized SHapley Additive exPlanations (SHAP) importance to predict the effect of spectral-shifting film of the logistic regression (LR), decision tree (DT), random forest (RF), XGBoost, and feedforward neural network (FFNN).}
    \label{fig:placeholder}
\end{figure}

\begin{figure}
    \centering
    \includegraphics[width=0.8\linewidth]{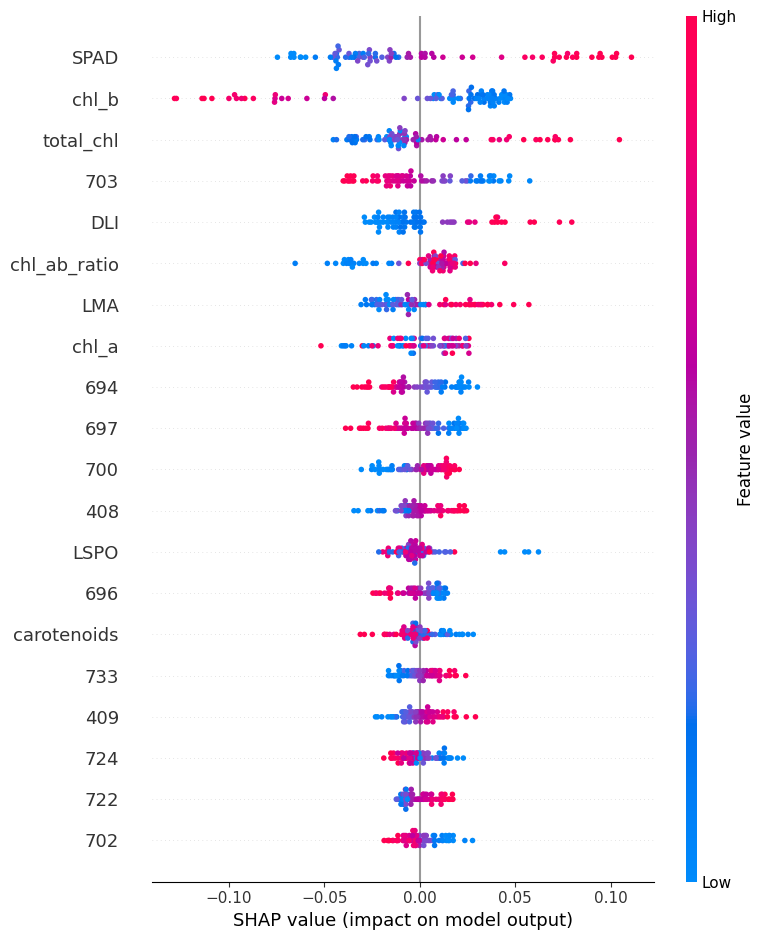}
    \caption{SHapley Additive exPlanations (SHAP) values for top 20 features in a feedforward neural network model predicting growth enhancement effects of spectral-shifting film.}
    \label{fig:placeholder}
\end{figure}

\section{Discussion}
\subsection*{Controversial growth enhancing effect of SF}
Empirical experiments conducted on 24 different crops
demonstrated the potential of SF to enhance crop productivity. However, the
modified solar spectrum through SF was not universally effective on plant
growth\cite{Paskhin2022}. Some crops exhibited significant yield increases
with SF, while other cultivars within the same crop did not respond similarly. For
example, among the lettuce cultivars tested, Jeock, Superseonpung, and Topgreen
showed significant increases of 19.8\%, 20.7\%, and 23.2\%, respectively, under SF
(Fig. 5). In contrast, other lettuce varieties such as Butterhead, Caesars
Green, Jeockchima, Multihead, Oak, and Yeolpung showed no difference in growth
or even decreased yields. A similar pattern was observed in Chinese cabbage. According
to previous study\cite{Kang2023a}, the green leaf cultivar (ACC202)
exhibited a significant increase of 38.3\% under SF, while the red leaf cultivar
(Jinhongssam) exhibited a 1.1\% reduction compared to PEF. These findings
underscore the necessity for a comprehensive understanding of the factors
influencing plant responses to SF, including phenotypic and environmental
traits. By elucidating these intricate relationships, the application of SF
technology can be optimized to maximize its benefits for diverse crop
production systems. In this study, the complexity of predicting the effect of
SF was addressed by employing machine learning and deep learning techniques.

\subsection*{Variational autoencoder effectively augment data}
Over the course of the three-year empirical study, 210
samples of growth results and phenotypic data for 24 different crops were
collected. Nevertheless, this amount of data is far from enough to train
machine learning and deep learning\cite{Feng2019}. To address this lack of
data, data augmentation with VAE was used\cite{chadebec2021}. VAE is a type of generative model that learns the probability
distribution of the given training data and generates new data points that
follow the same distribution\cite{kingma2022}. In this study, data
augmentation with VAE also represented the original data well (Fig. 6). Consistent with previous studies\cite{Huang2022, Saldanha2022, Izonin}, data augmentation improved model performance (Fig.
7). Deep learning models using artificial neural networks rely heavily on the
amount of data to train. This trend was evident in the increase
in test accuracy of FFNN with the number of training data.

\subsection*{Comparison of model performance for predicting the effect of SF}
Various machine learning and deep learning models were
evaluated for their performance to predict the effect of SF on crop yield (Fig.
7). All of the machine learning and deep learning models had significantly
higher validation accuracy, but not as high test accuracy. In particular, DT
had lower test accuracy, indicating overfitting. It is of paramount importance
to avoid overfitting, which is a poor generalization ability that performs well
on training data but poorly on test data. RF and XGBoost are ensemble models,
with several additional factors to avoid overfitting. RF limits the number of trees created and XGBoost numerically adjusts
the degree of regularization to reduce overfitting and increase the model performance.
The results of this study show that RF outperforms XGBoost. This is likely due
to differences in the fit of the data to the model, rather than differences in
the model performance\cite{bentejac2021}. The FFNN, trained on augment 4,
demonstrated superior performance compared to other models,
achieving an accuracy of 91.4\% on the test dataset. Crucially, the crops in the
test dataset were completely different from those used in training. This
highlights a significant advantage of the model: it can accurately identify the
growth enhancement effect of SF-modified sunlight on novel crops, based solely
on their phenotypic and photosynthetic traits along with daily light integrals.
This ability to generalize to unseen crops demonstrates the model's potential
as a powerful tool for predicting SF effectiveness across a wide range of plant
species without the need for extensive, species- and cultivar-specific
experimental trials. The ROC curves and AUC values further validated the strong
predictive power of FFNN, suggesting its potential for guiding SF application
in diverse crop production.

\subsection*{Analysis of key features influencing model prediction}
Feature importance provided insights into the key
phenotypic traits driving the model prediction\cite{Shin2021, Yoon2023, Yoon2024, Shin2024}. By extracting the SHAP importance for each
model, it was discovered that there are common features that predict the effect
of SF\cite{Lundberg2017}. LR and FFNN, which consider all features of the
data, determined that the SPAD and LMA, i.e. leaf thickness, were the most
important feature. This is consistent with previous research that suggests the
difference in growth enhancement between Chinese cabbage and lettuce grown in
SF may be due to leaf absorption, chl content, and leaf thickness\cite{Kang2023a}. Furthermore, it was determined that the concentrations of chl b, total
chl, and environmental traits (e.g. DLI and LSPO) were of greater importance
than spectral reflectance, except for the reflectance at 703 nm. The results of
the spectral SHAP importance analysis exhibited a similar pattern. In
particular, all models assessed the importance of reflectance around 700 nm in
determining the effectiveness of SF. The reflectance that SHAP determined to be
important was similar to the reflectance wavebands of chl a, b, and carotenoids\cite{Chazaux2022}. The optical properties of leaves are determined by the
interaction of chlorophyll composition, which varies among plants, and leaf
thickness. The interaction between internal leaf
phenotypic traits and light quality is not well understood. However, this study
provides novel evidence demonstrating that these traits are key determinants of
light treatment efficacy. Taken together, these results indirectly suggest that
the chlorophyll content and thickness of leaves play a crucial role in
determining the effectiveness of SF by increasing the absorption rate of
specific wavelengths. In other words, it implies that phenotypic
characteristics such as chlorophyll content and leaf thickness should be
considered to optimize the effect of SF.

Despite the promising results, this study has several
limitations that should be addressed in future research. First, the datasets
consisted of a limited number of crop species, while in real-world agriculture,
a much wider variety of crops are cultivated. Second, the effect of solar
spectrum modification through SF is to enhance photosynthetic capacity at light
intensities above the light saturation point\cite{Yoon2020, Kang2022}. Therefore, if solar radiation is limited due to seasonal variations, the
effectiveness of solar spectrum modifications may be limited, which can be seen
in the significantly higher importance of DLI and LSPO (Fig. 11). In addition, the phenotype of a plant can vary depending
on its growing environment, even for the same crop\cite{Coleman1994, Yan2013}. Consequently, the current dataset represents the phenotype in the
environment at the time of measurement, which may change as a result of
seasonal or climatic variations. These limitations suggest the need for further
experimentation in other environments.

\section{Conclusion}
This study presents a robust deep learning methodology to elucidate the unpredictable effects of SF on crop productivity. The core of this study lies in overcoming the limitations of the original dataset, which was absolutely insufficient for deep learning training, through VAE based data augmentation.The FFNN model trained on this augmented dataset demonstrated the ability to predict the yield enhancing effect of SF on untrained crops with a accuracy of 91.4\%. This study not only achieved high predictive performance as an engineering accomplishment but also scientifically demonstrated, through model interpretation using SHAP, that chlorophyll content, leaf refectance at 703 nm, DLI, and leaf thickness are the most crucial key predictors determining this complex biological response. In conclusion, this study provides an AI-based proof-of-concept that enables the precise determination of SF's efficacy without the need for costly and time-consuming real-world cultivation trials. This demonstrates that AI models can quantitatively capture species- and cultivar-specific responses in plants. This study opens a new pathway for establishing precise solar spectrum management strategies in agricultural practice.

\section{References}
\vspace{-1cm}
\renewcommand{\refname}{} 


\bibliographystyle{unsrt}  
\bibliography{ref}   
\end{document}